\documentclass{jpp}
\usepackage{graphicx,url,bm,natbib,color,macros}
\graphicspath{{./figs/}}
\usepackage{hyperref}

\title{Evolutionary tracks and isochrones of the inverse Hall cascade}

\shorttitle{Evolutionary tracks and isochrones of the inverse Hall cascade}
\shortauthor{Axel Brandenburg and Vinay Kumar}

\author{
Axel Brandenburg\aff{1,2,3,4}\corresp{\email{brandenb@nordita.org}}
and Vinay Kumar\aff{5,1}}

\affiliation{
\aff{1}Nordita, KTH Royal Institute of Technology and Stockholm University,\\
Hannes Alfv\'ens v\"ag 12, SE-10691 Stockholm, Sweden
\aff{2}The Oskar Klein Centre, Department of Astronomy,
Stockholm University, AlbaNova, SE-10691 Stockholm, Sweden
\aff{3}McWilliams Center for Cosmology \& Department of Physics,
Carnegie Mellon University, Pittsburgh, PA 15213, USA
\aff{4}School of Natural Sciences and Medicine, Ilia State University,\\
3-5 Cholokashvili Avenue, 0194 Tbilisi, Georgia
\aff{5}International Centre for Theoretical Sciences, Tata Institute of Fundamental Research, Bangalore 560089, India
}

\begin{document}

\maketitle

\begin{abstract}
Decaying MHD turbulence can exhibit an inverse cascade, which means that the height of the spectrum of magnetic energy in the subinertial range increases with time.
However, there is evidence for a resistive dependence of the inverse transfer speed.
It is unclear whether this is caused by reconnection, which is known to depend on resistivity.
Here, we show that in the Hall cascade, where no momentum equation is solved,
and the induction term is replaced by what corresponds to the negative Lorentz force in MHD,
there is still a resistive dependence of the inverse transfer speed,
but no such dependence is found for an analogous numerical reconnection experiment.
This suggests that the resistive dependence of the inverse transfer speed is not necessarily related to reconnection.
\end{abstract}

\section{Introduction}

Turbulent magnetic fields can exhibit an inverse cascade where the spectral magnetic energy increases with time at small wavenumbers.
There are many astrophysical systems that can exhibit inverse cascading of a turbulent magnetic field.
The large-scale magnetic fields of stars and galaxies are typical examples.
In two important papers, \cite{Frisch+75} and \cite{PFL76} found that the conservation of magnetic helicity can play an important role in driving an inverse cascade.
In the aforementioned examples, the magnetic fields are produced by dynamo action, which itself can produce magnetic fields at large length scales by what is known as the $\alpha$ effect \citep{SKR66}.
Decaying turbulence is another important area where the inverse cascade can manifest itself even more clearly.
The radiation-dominated era of the early universe is one of the most important applications \citep{BEO96, Christensson+01, BJ04}.

A completely different application is the magnetic field in neutron star crusts,
where the operation of an inverse cascade has been discussed \citep{Cho11, Bra20, Sarin+23, DB25}.
In the present paper, we focus on this case.
In such systems, only the electrons move, while the ions are considered immobile.
This is a good approximation for the crusts of neutron stars \citep{Goldreich+Reisenegger92}.
Therefore, the electron velocity $\vv_\mathrm{e}$ in the induction equation is proportional to the current density $\JJ$.
Specifically, we have $\vv_\mathrm{e}=-\JJ/e n_\mathrm{e}$ (in SI units),
where $e$ is the unit electric charge and $n_\mathrm{e}$ is the electron density.
Thus, the evolution of the magnetic field $\BB$ with time $t$ is governed by the equation
\begin{equation}
\frac{\partial\BB}{\partial t}=-\nab\times\left(\frac{1}{e n_\mathrm{e}}\JJ\times\BB+\eta\mu_0\JJ\right),
\label{dBdt}
\end{equation}
where the current density is related to the magnetic field through $\JJ=\nab\times\BB/\mu_0$,
$\eta$ is the magnetic diffusivity, and $\mu_0$ the magnetic permeability.
This equation is similar to the momentum equation \citep{Wareing+Hollerbach09, Wareing+Hollerbach10},
which motivated the expectation of a turbulent cascade \citep{Goldreich+Reisenegger92}.
In particular, there can be solutions that resemble turbulent cascades.
They were therefore called Hall cascades in subsequent papers \citep{Cumming+04, Wareing+Hollerbach09}.
Unlike ordinary magnetohydrodynamics, where one can think of the magnetic field as being characterized by an Alfv\'en speed with dimensions of $\m\s^{-1}$,
it is here characterized by the quantity $\Brms/e n_\mathrm{e}\mu_0$, which has dimensions of $\m^2\s^{-1}$ \citep[][hereafter B20]{Bra20}.
Looking at the two terms inside the parentheses in \Eq{dBdt}, this is not surprising, because the combination $\Brms/e n_\mathrm{e}\mu_0$ balances the
microphysical magnetic diffusivity $\eta$, which also has the dimensions $\m^2\s^{-1}$.
This difference in dimensions between Hall and MHD has significant consequences for the decay laws
of the magnetic field---even without net magnetic helicity \citep[][hereafter B23]{Bra23}.

In the present paper, we focus on the case of two-dimensional magnetic fields.
There is an important distinction between two-dimensional fields that are confined to the plane and those that have a component out of the plane.
To distinguish between these two cases, the latter is sometimes also referred to as 2.5-dimensional (2.5-D),
as opposed to ``strictly'' two-dimensional ones, when there is no component out of the plane.
This distinction is important, because in strictly two dimensions, the magnetic helicity density $\AAA\cdot\BB$ vanishes,
where $\AAA$ is the magnetic vector potential with $\BB=\nab\times\AAA$.

It is well known that the mean magnetic helicity density, $\HHM\equiv\bra{\AAA\cdot\BB}$, is conserved in MHD.
In strictly two dimensions, however, when the magnetic field lies entirely in the $xy$ plane, i.e., $\AAA\cdot\BB$ vanishes,
and the anastrophy $\bra{A_z^2}$, i.e., the mean squared normal component of $\AAA$, is conserved instead.

In the case of \Eq{dBdt}, we see that if the $\BB$ vectors were initially in the plane, $\JJ$ only has a component out of the plane.
Therefore, $\JJ\times\BB$ is, again, in the plane, and its curl is out of the plane---in contrast the original field, which was in the plane.
Therefore, in the Hall cascade, the magnetic field is necessarily 2.5-D.
Such magnetic fields could well be helical, but even when the net magnetic helicity vanishes, there are always magnetic helicity fluctuations.
In that case, the decay is governed by the conservation properties of the Hosking integral \citep{HS21,Zhou+22}.

\section{Dimensions in Hall and MHD}

The differences between the Hall and MHD cascades come from the fact that the natural dimension of the magnetic field in the Hall cascade
is $\m^2\s^{-1}$, while in MHD, it is $\m\s^{-1}$.
Using $e=1.6\times10^{-19}\A\s$, $\mu_0=4\pi\times10^{-7}\T\m\A^{-1}$, and
$n_\mathrm{e}\approx2.5\times10^{40}\m^{-3}$ for neutron star crusts at the surface \citep{Gourgouliatos+20}, we have
$en_\mathrm{e}\mu_0\approx5\times10^{15}\T\s\m^{-2}$, and therefore (B23)
\begin{equation}
\frac{B}{en_\mathrm{e}\mu_0}=\frac{B}{5\times10^{15}\T}\,\frac{\m^2}{\s}
=\frac{B}{1.6\times10^{12}\G}\,\frac{\km^2}{\Myr}.
\end{equation}
Without any further calculation, this equation allows us to state that, in a parametric representation of $B$ versus the typical magnetic length scale $\xiM$,
the magnetic field $B(t)$ should have reached a position $B(\xiM)=en_\mathrm{e}\mu_0\,\xiM(t)^2/t$ at time $t$ that is given by
\begin{equation}
B\left(\xiM(t)\right)=1.6\times10^{14}\G
\left(\frac{\xiM(t)}{10\km}\right)^2
\left(\frac{t}{1\Myr}\right)^{-1}.
\end{equation}
The concept of a parametric representation of $B$ vs.\ $\xiM$ has been borrowed from cosmology \citep{BJ04},
where a primordial magnetic field is undergoing a turbulent decay and will have reached
the termination line $B(\xiM)$ at the present time $t$ in a diagnostic diagram showing $B(t)$ vs.\ $\xiM(t)$.
At any earlier time, the line $B(\xiM)$ forms an isochrone for the time $t$.
The slope of the evolutionary track in such a diagram depends on the turbulent decay law, but the position of the isochrone does not.
However, the isochrones depend on the effective dimension of the magnetic field and have here the slope two,
which is different from ordinary MHD, where the slope is one \citep{BJ04}.

The full range of differences between the inverse MHD and Hall cascades is rather complex and therefore best illustrated in tabular form; see \Tab{TScalings0}.
We begin by summarizing the basic dimensions of the mean squared magnetic field $\EEM(t)=\bra{(B/en_\mathrm{e}\mu_0)^2}/2$,
which in MHD corresponds to the mean magnetic energy per unit mass.
Next, we summarize the functional dependence of $\EEM(t)$ on $\xiM$ and $t$.
The actual termination line $B(\xiM)$ may occur for a time that is shorter by a factor $C_\mathrm{M}$.
This factor can be determined by plotting the time evolution of
\begin{equation}
\CM^\mathrm{MHD}(t)=\EEM^{1/2}\,t\;\xiM^{-1}
\end{equation}
in MHD and, correspondingly,
\begin{equation}
\CM^\mathrm{Hall}(t)=\EEM^{1/2}\,t\;\xiM^{-2}
\end{equation}
for the Hall cascade.

\begin{table}\caption{
Scalings for Hall and MHD cascades in general.
}\vspace{12pt}\centerline{\begin{tabular}{lccc}
                  &           & MHD                        & Hall \\
\hline
$[\EEM]$          & $=$       & $\m^2\s^{-2}$              & $\m^4\s^{-2}$ \\
$[\HHM]$          & $=$       & $\m^3\s^{-2}$              & $\m^5\s^{-2}$ \\
$\EEM(t)$         & $\propto$ & $\xiM(t)^2 \, t^{-2}$      & $\xiM(t)^4 \, t^{-2}$ \\
$C_\mathrm{M}(t)$ & $=$       & $\EEM^{1/2}\,t\,\xiM^{-1}$ & $\EEM^{1/2}\,t\,\xiM^{-2}$ \\
\label{TScalings0}\end{tabular}}\end{table}

With these preparations, we can now summarize the mathematical expressions for different
conserved quantities: the mean magnetic helicity density $I_\mathrm{M}$ for helical fields,
the Hosking integral in nonhelical turbulence (in 3-D and 2.5-D), and the anastrophy $I_\mathrm{A}$ in
the strictly 2-D case where the magnetic field has no component out of the plane.
Next, we summarize the scalings of $\xiM$ and $\EEM$, the parametric representation of $B(\xiM)$,
and the slope of the outer envelope of the peak of the spectra $\max[E(k,t)]$ vs.\ $k$.

\begin{table}\caption{
Scalings for inverse Hall and MHD cascades.
Square brackets indicate that Hall cannot be realized in 2-D because of the out-of-plane component.
}\vspace{12pt}\centerline{\begin{tabular}{lllll}
Case         & Quantity         & MHD                   & Hall & reference \\
\hline
helical      & $[I_\mathrm{M}]$ & $\propto \m^3\s^{-2}$ & $\propto \m^5\s^{-2}$ & section~2.7 of B20 \\
(for 3-D     & $\xiM$           & $\propto t^{2/3}$     & $\propto t^{2/5}$    & figure~6 of B20 \\
or 2.5-D)    & $\EEM$           & $\propto t^{-2/3}$    & $\propto t^{-2/5}$   & figure~6 of B20 \\
             & $\Brms$          & $\propto\xiM^{-1/2}$  & $\propto\xiM^{-1/2}$ & equation~(27) of B20 \\
             & $E(k,t)_{\max}$  & $\propto k^{0}$       & $\propto k^{0}$ & figure~7 of B20 
\vspace{5pt}\\
3-D nonhel   & $[I_\mathrm{H}]$ & $\propto \m^9\s^{-4}$ & $\propto \m^{13}\s^{-4}$ & equation~(2.4) of B23 \\
             & $\xiM$           & $\propto t^{4/9}$     & $\propto t^{4/13}$   & figure~4 of B20 \\
             & $\EEM$           & $\propto t^{-10/9}$   & $\propto t^{-10/13}$ & figure~1 of B23 \\
             & $\Brms$          & $\propto\xiM^{-5/4}$  & $\propto\xiM^{-5/4}$ & --- \\
             & $E(k,t)_{\max}$  & $\propto k^{3/2}$     & $\propto k^{3/2}$    & figure~3 of B20
\vspace{5pt}\\
2.5-D nonhel & $[I_\mathrm{H}]$ & $\propto \m^8\s^{-4}$ & $\propto \m^{12}\s^{-4}$ & present work \\
             & $\xiM$           & $\propto t^{1/2}$     & $\propto t^{1/3}$  & \Fig{pkpm_2048_k1_e1em8_k60a_h2em13} of present work \\
             & $\EEM$           & $\propto t^{-1}$      & $\propto t^{-2/3}$ & \Fig{pkpm_2048_k1_e1em8_k60a_h2em13} of present work \\
             & $\Brms$          & $\propto\xiM^{-1}$    & $\propto\xiM^{-1}$ & \Fig{pkpm_2048_k1_e1em8_k60a_h2em13} of present work \\
             & $E(k,t)_{\max}$  & $\propto k^{1}$       & $\propto k^{1}$    & \Fig{ppower_comp} of present work
\vspace{5pt}\\
2-D nonhel   & $[I_\mathrm{A}]$ & $\propto \m^4\s^{-2}$ &[$\propto \m^6\s^{-2}$] & not conserved for Hall! \\
             & $\xiM$           & $\propto t^{1/2}$     &[$\propto t^{1/3}$]     & --- \\
             & $\EEM$           & $\propto t^{-1}$      &[$\propto t^{-2/3}$]    & --- \\
             & $\Brms$          & $\propto\xiM^{-1}$    &[$\propto\xiM^{-1}$]    & --- \\
             & $E(k,t)_{\max}$  & $\propto k^{1}$       &[$\propto k^{1}$]       & --- \\
\label{TScalings}\end{tabular}}\end{table}
In MHD, many of the expressions summarized here have previously been verified numerically and theoretically.
For numerical work trying to determine $C_\mathrm{M}^\textrm{MHD}$, we refer to \cite{BNV24}.
For the Hall cascade, however, $C_\mathrm{M}^\textrm{Hall}$ has not yet been determined.

\section{Our model}

\subsection{Initial conditions}

Our initial magnetic vector potential is given by
\begin{equation}
A_i({\kk})=A_0\left[P_{ij}(\kk)-\ii\sigma\epsilon_{ijl}
k_l/k\right] g_j({\kk})\, S(k),
\label{Bikk}
\end{equation}
where $g_j(\kk)$ is the Fourier transform of a $\delta$-correlated vector field with Gaussian fluctuations,
$k_0$ is the initial wavenumber of the energy-carrying eddies, $\sigma$ is the magnetic helicity, and $S(k)$ determines the spectral shape with (B20)
\begin{equation}
S(k)={k_0^{-1/2} (k/k_0)^{\alpha/2-1}
\over[1+(k/k_0)^{2(\alpha+7/3)}]^{1/4}},
\label{Sfunction}
\end{equation}
where $\alpha$ is the subinertial range slope.
We apply periodic boundary conditions in all directions.
We use the \textsc{Pencil Code} \citep{PC} for our simulations, where we solve for the magnetic vector potential $\AAA$ such that $\BB=\nab\times\AAA$ is always solenoidal.

In our numerical calculations, we always set the domain size to $L=2\pi$, i.e., the lowest wavenumber in the domain is $k_1=2\pi/L=1$.
We also set $en_\mathrm{e}=\mu_0=1$, i.e., the Hall parameter is always set to unity.
This fixes the dimensions of all other quantities.
In the following, we vary $k_0$, $A_0$, $\eta$, $\eta_\mathrm{hyp}$, and the number of mesh points.
Except for Run~N (see \App{HelicalDecay}), where $\sigma=1$, we set $\sigma=0$ in all other cases.
Likewise, except for one case with $\alpha=4$, we always use $\alpha=3$.

\subsection{Diagnostic quantities}
\label{DiagnosticQuantities}

We define the magnetic energy spectrum $\EM(k,t)$ such that $\int\EM(k,t)\,\dd k=\bra{\BB^2/2\mu_0}$.
Furthermore, the integral scale of the magnetic field is defined as $\xiM=\int k^{-1}\EM(k,t)\,\dd k/\EEM$.
In such a Hall cascade, where ions do not play a role, we can define a Lundquist number as $\mathrm{Lu}(t)=\Brms/en_\mathrm{e}\mu_0\eta$ (B20).
We also define the instantaneous scaling exponents $p(t)=-\dd\ln\EEM/\dd\ln t$ and $q(t)=\dd\ln\xiM/\dd\ln t$.

\subsection{Possibility of using magnetic hyperdiffusivity}

In MHD, the magnetic diffusion term enters with a higher number of derivatives than the nonlinear term.
Therefore, magnetic diffusivity always dominates at small length scales.
This is different for the Hall cascade.
Indeed, in \Eq{dBdt}, both the magnetic diffusion term and the nonlinear term are proportional to $\JJ$, i.e., they enter with the same number of derivatives of the magnetic field.
Therefore, discretization errors cannot efficiently be removed by the magnetic diffusion operator \citep{Wareing+Hollerbach10,Cho11}.
This can lead to oscillations (or ``ringing'') on the mesh scale.

To suppress the effect of discretization errors on the mesh scale, we can add magnetic hyperdiffusivity,
i.e., a magnetic diffusion operator with a higher number of derivatives; see, e.g., \cite{BM99}.
In the \textsc{Pencil Code}, we can use a diffusion operator proportional to $\nabla^6$, so the uncurled induction equation becomes
\begin{equation}
\frac{\partial\AAA}{\partial t}=-\JJ\times\BB+\eta\nabla^2\AAA+\eta_\mathrm{hyp}\nabla^6\AAA,
\label{dAdt}
\end{equation}
where $\eta_\mathrm{hyp}$ is the magnetic hyperdiffusivity.

\begin{table}\caption{
Summary of our runs.
}\vspace{12pt}\centerline{\begin{tabular}{ccrccccc}
Run & $\alpha$ & $k_0$ & $A_0$ & $\eta$ & $\eta_\mathrm{hyp}$ & $\sigma$ & $N^2$ \\
\hline
A & 4 & 180 & 0.5 & $10^{-5}$ &       0           & 0 & $2048^2$ \\
B & 3 & 180 & 0.5 & $10^{-5}$ &       0           & 0 & $2048^2$ \\
C & 3 &  60 & 0.2 & $10^{-9}$ &       0           & 0 & $4096^2$ \\
D & 3 &  60 &  1  & $10^{-8}$ & $2\times10^{-13}$ & 0 & $2048^2$ \\
E & 3 &  60 &  1  & $10^{-8}$ & $       10^{-14}$ & 0 & $2048^2$ \\
F & 3 &  60 &  1  & $10^{-7}$ & $       10^{-14}$ & 0 & $2048^2$ \\
G & 3 &  60 &  1  & $10^{-6}$ & $       10^{-14}$ & 0 & $2048^2$ \\
H & 3 & 180 &  1  & $10^{-6}$ & $       10^{-14}$ & 0 & $2048^2$ \\
I & 3 & 180 &  3  & $10^{-6}$ & $       10^{-14}$ & 0 & $2048^2$ \\
J & 3 &  60 & 0.2 & $10^{-6}$ & $       10^{-14}$ & 0 & $2048^2$ \\
K & 3 &  60 &  1  & $10^{-5}$ & $       10^{-14}$ & 0 & $2048^2$ \\
L & 3 &  60 &  1  & $10^{-4}$ & $       10^{-14}$ & 0 & $2048^2$ \\
M & 3 &  60 &  1  & $10^{-3}$ & $2\times10^{-13}$ & 0 & $2048^2$ \\
N & 3 &  60 &  1  & $10^{-6}$ & $2\times10^{-13}$ & 1 & $2048^2$ \\
\label{TSummary}\end{tabular}}\end{table}

\section{Results}

\subsection{Subinertial range spectrum}

\begin{figure}\begin{center}
\includegraphics[width=\columnwidth]{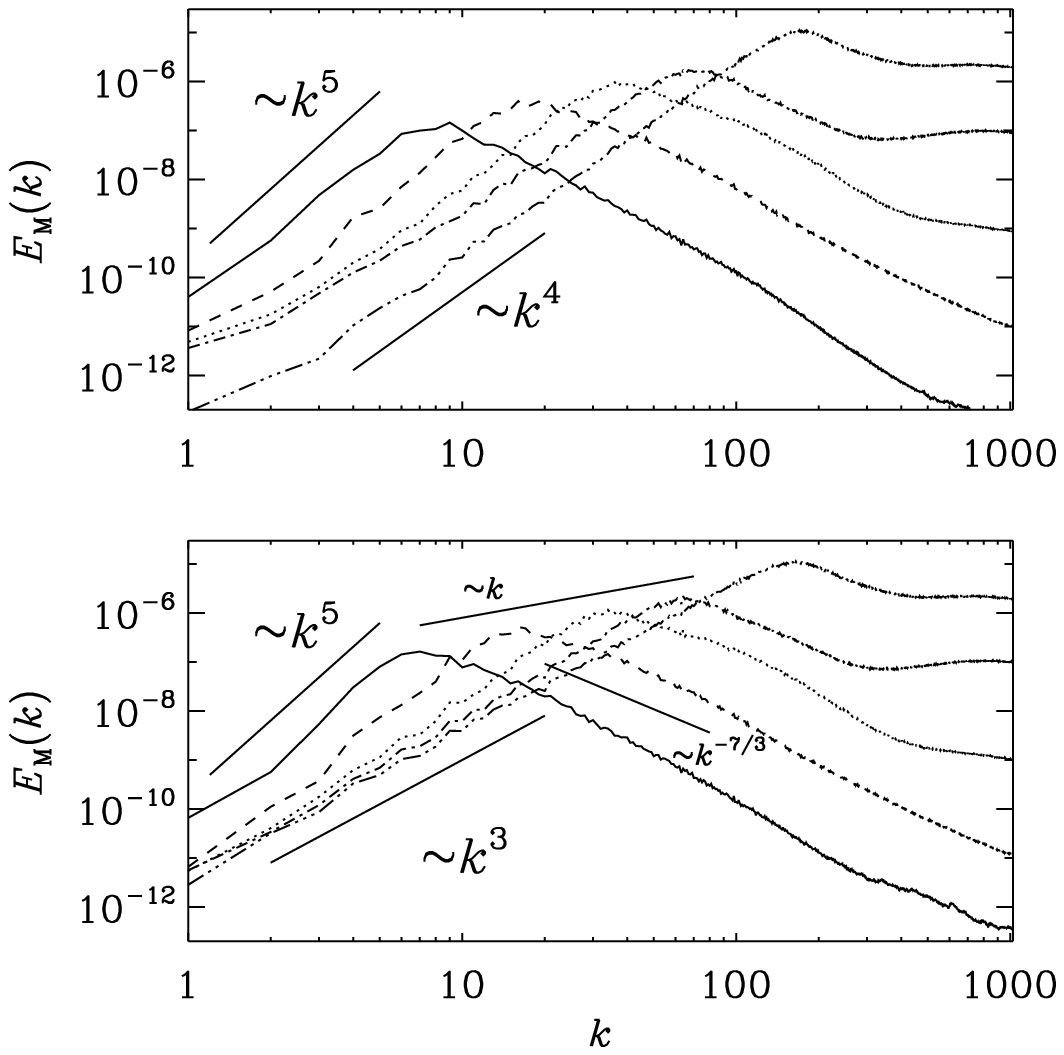}
\end{center}\caption[]{
Spectral evolution for Runs~A and B with initial (a) $\propto k^4$ and (b) $\propto k^3$ spectra, respectively.
The times are $t=0.001$, 0.1, 1, 10, and 100.
}\label{ppower_comp}\end{figure}

A natural initial condition for the magnetic vector potential at large scales would be a random magnetic vector potential.
It is therefore also called a causal spectrum \citep{Durrer+98, DC03}, which is routinely used as initial condition in cosmological applications \citep{Christensson+01}.
In 3-D, this corresponds to an initial $k^4$ spectrum for the shell-integrated magnetic energy spectrum, but in 2-D, it corresponds to an initial $k^3$ spectrum.
It turns out that both initial conditions lead to inverse cascading with a spectrum that becomes steeper than $k^4$ and is closer to $k^5$; see \Fig{ppower_comp}.
This was also already found in 3-D; see figure~3 of B20.
At early times, on the other hand, an initial $k^3$ spectrum leads to a slower development of the inverse cascade.

We also see that the peaks of the spectra evolve underneath a $k^\beta$ envelope with $\beta=1$,
which is shallower than what is obtained in 3-D, where $\beta=3/2$ (see figure~3 of B20).
We also see that there is an inertial range with a $k^{-7/3}$ spectrum \citep{Biskamp+96}.
In this context, it is interesting to note that \cite{Wareing+Hollerbach09b} proposed a slightly steeper $k^{-5/2}$ spectrum.

\subsection{Visual appearance of the Hall cascade}

\begin{figure}\begin{center}
\includegraphics[width=\columnwidth]{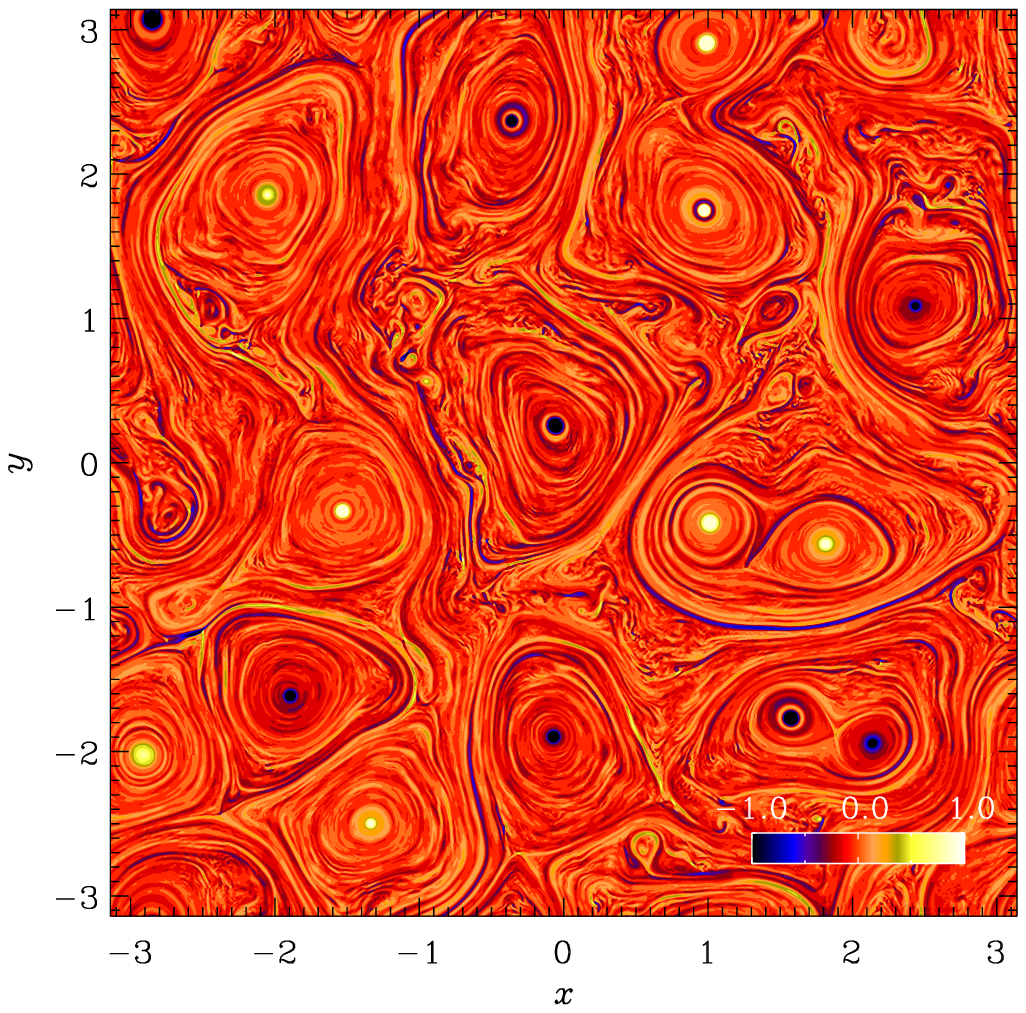}
\end{center}\caption[]{
Visualization of $J_z(x,y)$ for Run~D at $t=100$.
}\label{pj_2048_k1_e1em8_k60a_h2em13}\end{figure}

We have seen in \Fig{ppower_comp} that in the 2.5-D Hall cascade, the inertial range has the same $k^{-7/3}$ spectrum as in the 3-D Hall cascade.
It is slightly steeper than for 3-D Kolmogorov-type turbulence in MHD \citep[see also][]{GS95}.
To get an idea about the spatial distribution of scales, we show in \Fig{pj_2048_k1_e1em8_k60a_h2em13} a visualization of $J_z(x,y)$.
The current density is dominated by circular islands of positive and negative sign, which are surrounded by layers of thin current sheets.
At the end of the simulation, the current sheets can extend over a major fraction of the computational domain.
The vertices between the convex current sheets are filled with much smaller scale current structures.
These small-scale structures are reminiscent of what can be called plasmoids in MHD; see, e.g., figure~14 of \cite{BNV24}.

\subsection{Scaling exponents in the inverse Hall cascade}

\begin{figure}\begin{center}
\includegraphics[width=\columnwidth]{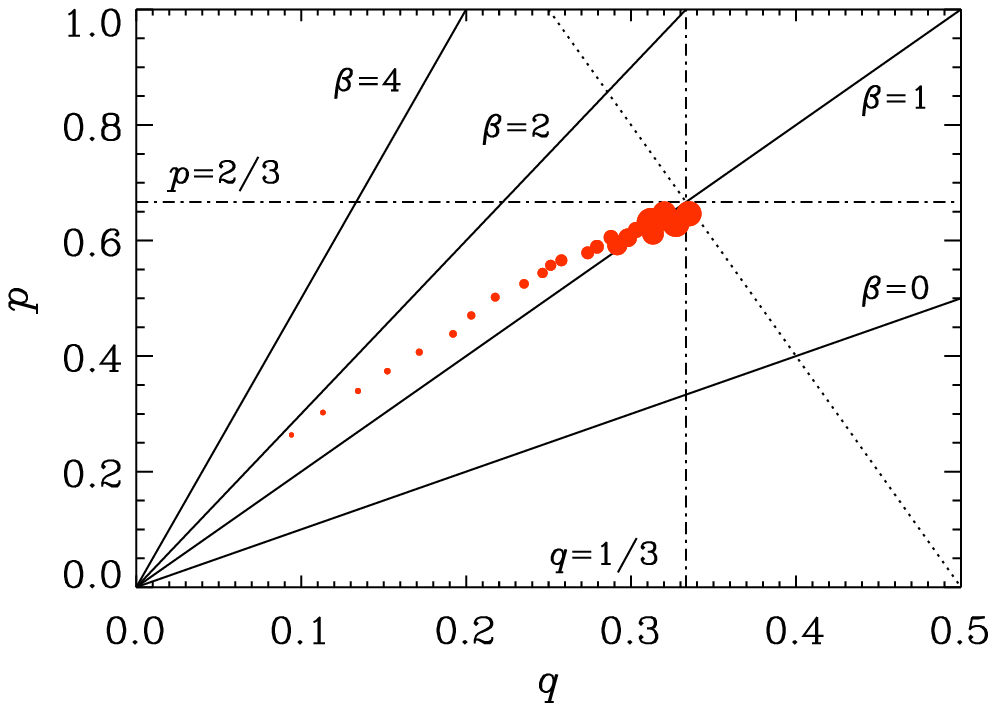}
\end{center}\caption[]{
$pq$ diagram for Run~D.
Larger symbols correspond to later times.
The solid lines show $\beta=0$, 1, 2, and 4, while the dashed-dotted lines indicate the positions $p=2/3$ and $q=1/3$.
}\label{pkpm_2048_k1_e1em8_k60a_h2em13}\end{figure}

The inverse Hall cascade is characterized by the instantaneous scaling exponents $p(t)$ and $q(t)$,
as well as the slope $\beta=p/q-1$ of the envelope underneath the evolving spectral peaks.
Similarly to the MHD case, it is instructive to plot $p(t)$ vs.\ $q(t)$ \citep{BK17}.
As time goes on, the points $(q,p)$ tend to move along a line $\beta=\const$.
In the present case, these points are initially closer to the $\beta=3$ line, but then approach the $\beta=1$ line; see \Fig{pkpm_2048_k1_e1em8_k60a_h2em13}.
Eventually, they stop on what can be called the Hall line, $p=2(1-2q)$, in analogy to the Alfv\'en line $p=2(1-q)$ in MHD; see B20 for details.

\begin{figure}\begin{center}
\includegraphics[width=\columnwidth]{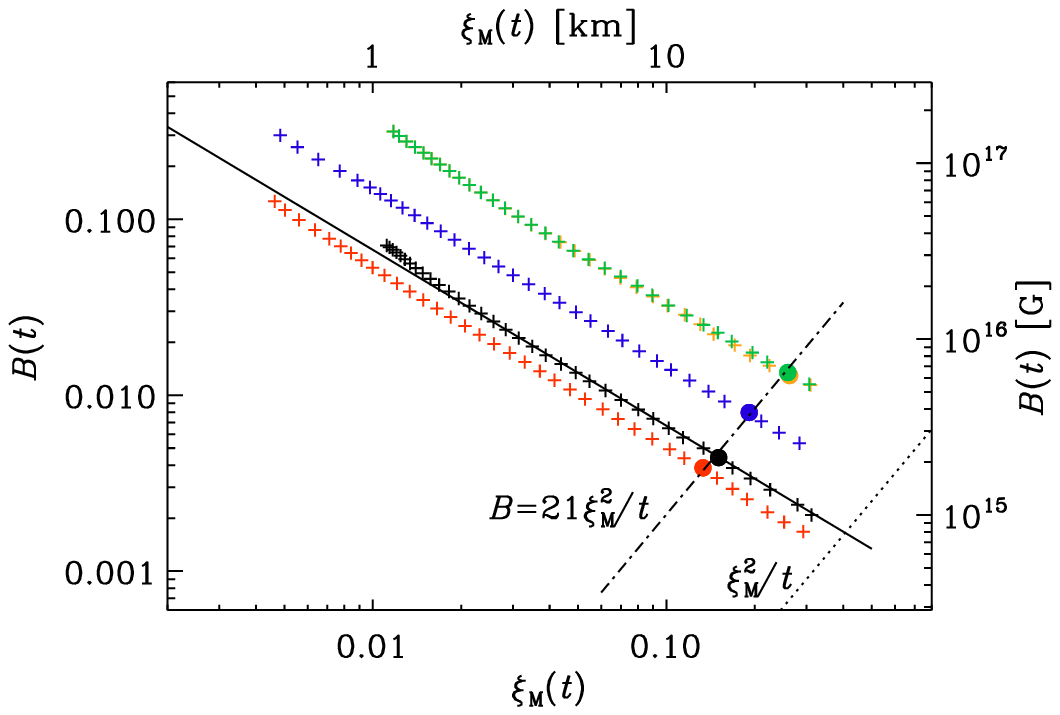}
\end{center}\caption[]{
Evolutionary tracks for Runs~E (green), G (orange), H (red), I (blue), and J (black).
The plus signs indicate times in logarithmic intervals of one sixth per decade.
The dashed-dotted line indicates the isochrone $B\approx\CM\xiM^2/t$ for the time $t=100$.
The filled symbols denote the data points for the same time.
The dotted line shows the corresponding Hall line for which $B=\xiM^2/t$.
}\label{pkpm_vs_lam}\end{figure}

\subsection{Evolutionary tracks and isochrones in the diagnostic diagram}

In the cosmological context, the diagram showing $B(t)$ vs.\ $\xiM(t)$ is sometimes referred to as diagnostic diagram,
because both quantities are observationally constrained, and certain regions in the diagram can then be excluded.
As already indicated in \Tab{TScalings}, the evolutionary tracks follow a $B\propto\xiM^{-1}$ law both in the inverse Hall cascade and in MHD.
Mathematically, it is indistinguishable from the strictly 2-D case, although $I_\mathrm{A}$ is not conserved in 2.5-D.

\FFig{pkpm_vs_lam} shows a diagnostic diagram for different values of $A_0$, $k_0$, and $\eta$.
The symbols show nearly parallel evolutionary tracks.
The endpoints of the evolution lie on a line $\propto\xiM^2/t$ for the Hall cascade, compared to a $\propto\xiM/t$ line for MHD.
This is well borne out by the simulations.
However, as in MHD, the endpoints lie by a certain amount shifted to the upper left, so the actual line is therefore $B\approx\CM\xiM^2/t$.

To compute the factor by which the endpoints are shifted to the upper left in the $B(t)$ vs.\ $\xiM(t)$ diagram,
it is convenient to plot the time-dependent function
\begin{equation}
C_\mathrm{M}(t)=\frac{\Brms}{e n_\mathrm{e}\mu_0}\,\frac{t}{\xiM^2(t)}
\end{equation}
and determine $\CM\to\lim_{t\to\infty} C_\mathrm{M}(t)$, where this limit is taken subject to times
when the correlation length is still contained within the finite computational domain.
\FFig{pkpm_vs_lam} also shows that for smaller values of $\eta$, with green being the lowest value ($10^{-8}$),
the endpoints are nearly exactly the same, and the actual trajectory is also almost exactly the same.

\begin{figure}\begin{center}
\includegraphics[width=\columnwidth]{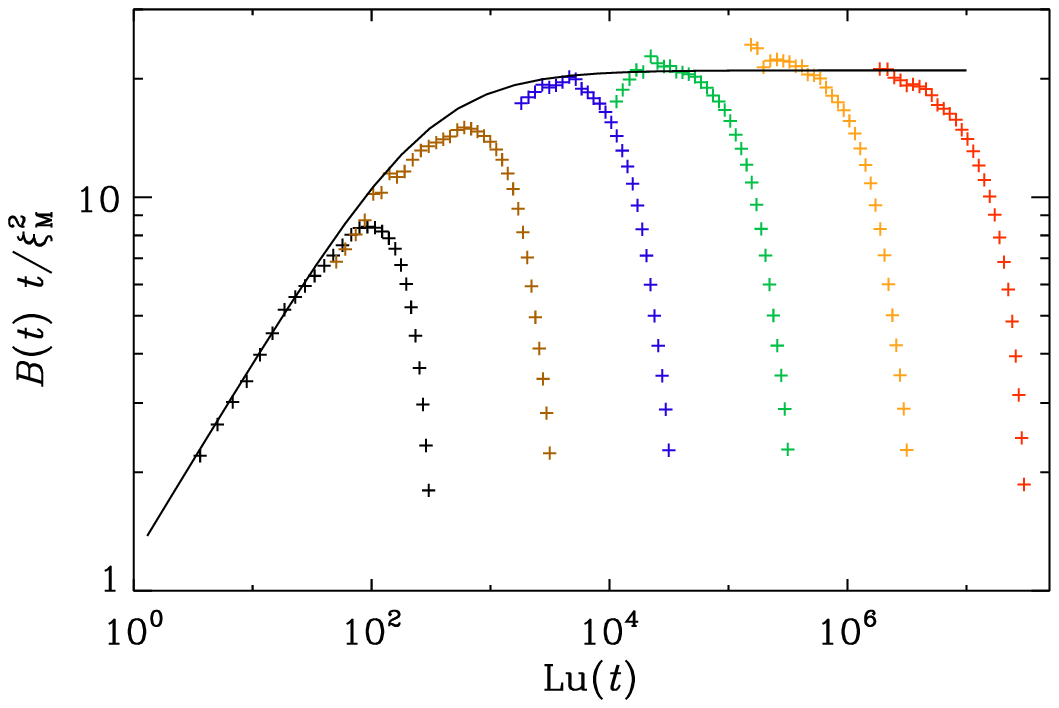}
\end{center}\caption[]{
Parametric representation of $\CM(t)$ vs.\ $\Lu(t)$ for Runs~D (red), F (orange), G (green), K (blue), L (brown), and M (black).
}\label{pkpm_comp_Lu}\end{figure}

\FFig{pkpm_comp_Lu} shows that for small values of $\Lu(t)$,
the value of $\CM(t)$ increases like $\Lu^{1/2}$, but for
$\Lu>\Lu_0\approx4\times10^2$, it saturates at $C_\mathrm{M}^{\max}\approx21$.
We overplot the curve
\begin{equation}
C_\mathrm{M}\approx C_\mathrm{M}^{\max}\,\Lu^{1/2}\left/\sqrt{\Lu_0+\Lu},\right.
\label{CM}
\end{equation}
with $C_\mathrm{M}^{\max}\approx21$ and $\Lu_0\approx3\times10^2$.
It fits the data reasonably well for not too large times.
This is because eventually the peak of the spectrum reaches the lower cutoff value $k=k_1$.

\begin{figure}\begin{center}
\includegraphics[width=\columnwidth]{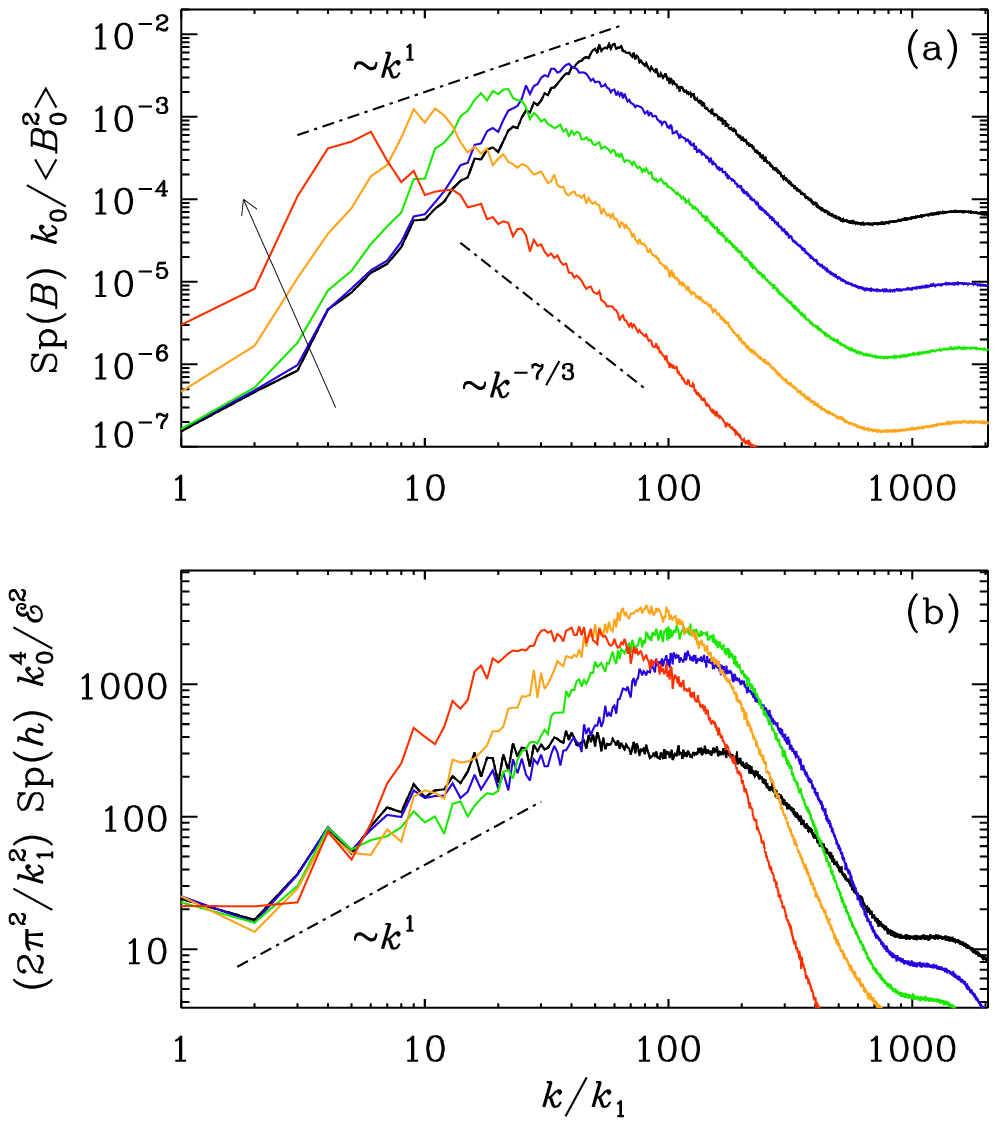}
\end{center}\caption[]{
Evolution of (a) $\Sp(B)$ and (b) $\Sp(h)$ for Run~C.
}\label{rspec_select_hoskM}\end{figure}

\begin{figure}\begin{center}
\includegraphics[width=\columnwidth]{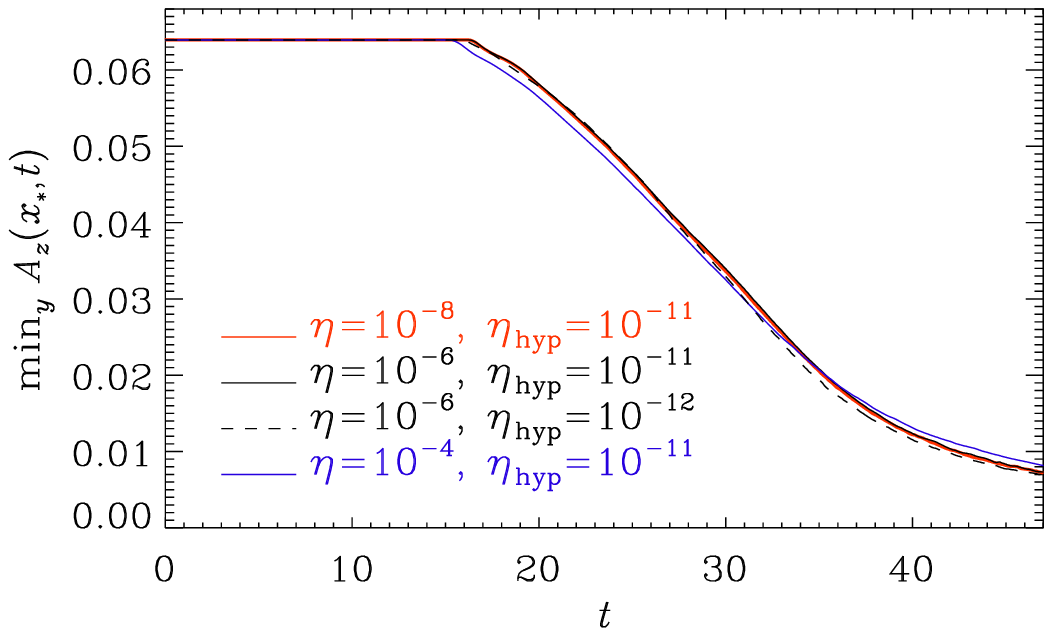}
\end{center}\caption[]{
Evolution of $\min_y A_z(x_\ast,t)$ for $\eta=10^{-8}$ (red), $10^{-6}$ (black), and $10^{-4}$ (blue) using $\eta_\mathrm{hyp}=10^{-11}$.
For $\eta=10^{-6}$, we also compare with $\eta_\mathrm{hyp}=10^{-12}$ (black dashed line).
}\label{pcomp_TMI}\end{figure}

\begin{figure}\begin{center}
\includegraphics[width=\columnwidth]{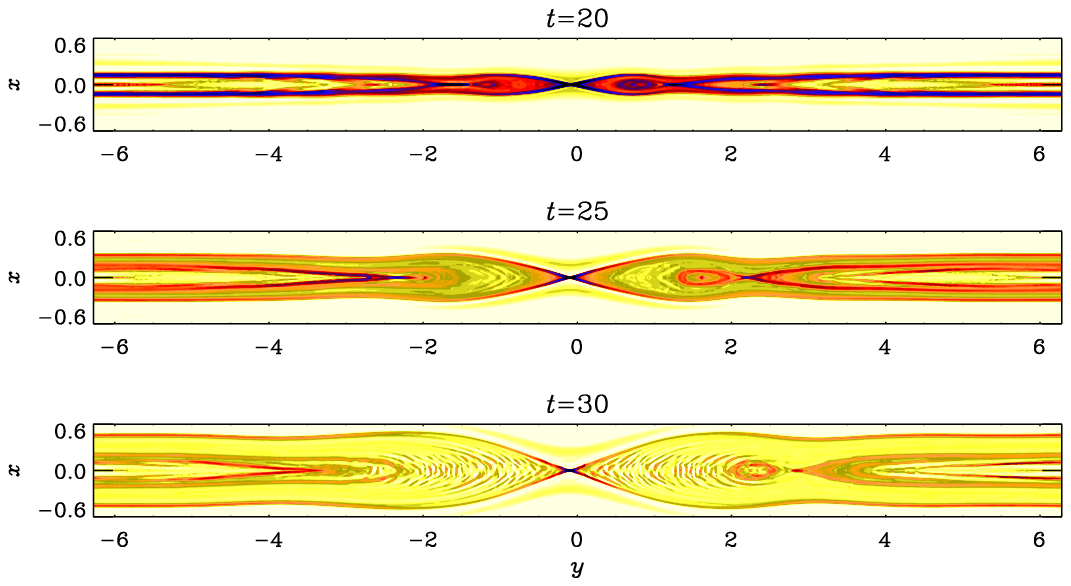}
\end{center}\caption[]{
Visualization of the current density for the run with $\eta=10^{-6}$ and $\eta_\mathrm{hyp}=10^{-11}$ at $t=20$, 25, and 30.
Light shades indicate small positive values and dark shades indicate large positive values.
}\label{pj_comp_2048_e1em6a_h1e11_rep}\end{figure}

\subsection{Conservation of the Hosking integral}

The Hosking integral quantifies the value of the magnetic helicity variance spectrum in the limit of small wavenumber, i.e., $\Sp(h)$ for $k\to0$, where $h=\AAA\cdot\BB$.
This follows from the Taylor expansion of $\Sp(h)=I_\mathrm{H}k^1/2\pi^2+...$; cf.\ with \cite{HS21} for the 3D case, where the spectrum is $\propto k^2$.
From \Fig{rspec_select_hoskM}, we see that $\Sp(h)$ is indeed nearly invariant in time.
By contrast, the magnetic flux correlation, i.e., the magnetic energy spectrum $\EM(k)=\sp(\BB)$ at small $k$, is not conserved.

\subsection{Comparison with reconnection}

It is well known that in MHD, the Hall term facilitates fast reconnection; see \cite{Zweibel2009} for a review.
In the following, we present 2-D results for our present system without momentum equation.
As initial condition, we adopt a current sheet given by
\begin{equation}
A_z=\frac{3\sqrt{3}}{4}\,B_0\,\delta\,{\rm sech}^2(x/\delta),
\label{sech-profile}
\end{equation}
where the prefactor $3\sqrt{3}/4\approx1.3$ ensures that $\max|B_y|=B_0$.
We use $B_0=0.1$ and $\delta=0.5$.
Unlike the Harris sheet, where $B_y\propto\tanh(x/\delta)$, the field
$B_y=-\partial_x A_z$ vanishes away from the sheet and the layer carries
no net current, so the profile is compatible with periodic boundaries.
It is tearing unstable for $k\delta<\sqrt{5}$.
With $en_\mathrm{e}\mu_0=1$, the Lundquist number based on the peak field is
$\Lu=B_0/en_\mathrm{e}\mu_0\eta=10^3$, $10^5$, and $10^7$ for
$\eta=10^{-4}$, $10^{-6}$, and $10^{-8}$, respectively.
To break the otherwise perfect symmetry, we added Gaussian distributed noise to $A_z$.

The reconnection speed can be determined as the negative slope of the minimum value
of $A_z(x_\ast,y,t)$ over $y$ along the current sheet at $x_\ast=0$.
In \Fig{pcomp_TMI}, we plot this as a function of $t$ for three values of the magnetic
diffusivity ($\eta=10^{-8}$, $10^{-6}$, and $10^{-4}$) using $\eta_\mathrm{hyp}=10^{-11}$.
For $\eta=10^{-6}$, we also compare with $\eta_\mathrm{hyp}=10^{-12}$.

It turns out that the results are fairly insensitive to the value of $\eta$.
However, for $\eta=10^{-4}$, corresponding to $\Lu=10^3$, the slope of the minimum value of $A_z(x_\ast,y,t)$ is slightly shallower.
The results are also fairly insensitive to the value of $\eta_\mathrm{hyp}$,
except that at late times, the function $A_z(x_\ast,y,t)$ drops slightly deeper for $\eta_\mathrm{hyp}=10^{-12}$.

Unlike the cases of turbulent decay in the Hall cascade, where $\EEM(t)$ and $\xiM(t)$, and therefore also $\CM(t)$, are insensitive to the value of $\eta_\mathrm{hyp}$,
this is not the case for Hall reconnection, where the reconnection speed is more sensitive to the magnetic field evolution at small length scales.
Note also that visualizations of $J_z(x,y)$ show a clear \textsc{X}-point without a flat current sheet, 
unlike in resistive MHD; see \Fig{pj_comp_2048_e1em6a_h1e11_rep}.
This is reminiscent of fast reconnection of \cite{Petschek64} type; see \cite{Zweibel2009}.

\section{Conclusions}

In 2.5-D, the inverse Hall cascade is qualitatively similar to that in 3-D.
The spectral energy at the smallest wavenumber increases with time both in the helical and nonhelical cases.
The scaling exponents are different in 2.5-D, but they follow the theoretical predictions: $p=2/3$ instead of 10/13, $q=1/3$ instead of $4/13$, and $\beta=1$ instead of 3/2.
Moreover, the isochrones (endpoints of the tracks) are shifted away from the Hall line by a resistivity-dependent factor $\CM$.
Similarly in MHD, the isochrones are shifted away from the Alfv\'en line by a resistivity-dependent factor.
In the MHD case, it has been argued that the resistivity dependence can be explained by reconnection \citep{HS23}.
In the present case, however, this explanation is ruled out, because Hall reconnection is not resistivity dependent.

In MHD, there is great significance in the question of the resistivity dependence of the isochrones.
Especially the endpoints of the evolution, i.e., the isochrones belonging to the present time, are conceptually important in cosmology, because we need to extrapolate over many orders of magnitude.
This can only be done with the correct theory.
Whether or not the position of the endpoints has to do with reconnection or with some other resistivity-dependent mechanism such as local magnetic helicity conservation \citep{BNV24}
has implications for our extrapolation.
Reconnection, for example, may depend on the value of the magnetic Prandtl number \citep{Comisso+15}, while local magnetic helicity conservation does not.

\cite{KB26} have argued that the magnetic-Prandtl-number dependence found by \cite{Comisso+15} could be a consequence of their usage of driven reconnection.
It is also possible that their runs were not yet in the deeply plasmoid-mediated regime.
By contrast, the results of \cite{KB26} for spontaneous reconnection are in the deeply plasmoid-mediated regime and do not show magnetic-Prandtl-number dependence.
Their finding could potentially alleviate the discrepancies found between reconnection and the endpoints in turbulent MHD evolution.
However, our conclusions regarding the role of reconnection could be affected by numerical resolution \citep{Anandavijayan+Bhat26}.

While our present work sheds some light on the aforementioned problem, it could also have implications on the magnetic field decay in neutron star crusts.
In the context of neutron stars, \Fig{pkpm_vs_lam} provides a novel framework for the analysis of magnetic field strengths and characteristic length scales.
In particular, this line of endpoints suggests that magnetic fields above $10^{15}\G$ can only be compatible
with a turbulent evolution if the length scale exceeds the length scale of the neutron star itself.
Whether or not this is indeed the case, and how this conclusion depends on our estimates entering this conclusion, still needs to be analyzed in more detail.

\section*{Acknowledgements}
The simulations were carried out on the \emph{Dardel} supercomputing cluster under allocations granted by the Swedish National Infrastructure for Computing through the Center for Parallel Computers at the Royal Institute of Technology (KTH), Stockholm.
V.K.\ gratefully acknowledges the hospitality and support of Nordita through the Nordita Visiting PhD Fellowship, as well as support from the Department of Atomic Energy, Government of India, under Project Nos.\ RTI4013 and RTI4019.
A.B.\ was supported in part by the European Research Council through the ERC Synergy Grant COSMOMAG under grant No.\ 101224803,
the Swedish Research Council (Vetenskapsr{\aa}det) under grant No.\ 2025-05957,
the National Science Foundation under grant Nos.\ NSF AST-2307698, AST-2408411, and NASA Award 80NSSC22K0825.

\section*{Data availability statement}

The data that support the findings of this study are openly available on
Zenodo at \href{https://doi.org/10.5281/zenodo.23697852}{doi:10.5281/zenodo.23697852} (v2026.10.27) or, for easier access, at:\\
\url{http://norlx65.nordita.org/~brandenb/projects/Hall-Endpoints/}.\\
All calculations have been performed with the {\sc Pencil Code}
\citep{PC};
\href{https://doi.org/10.5281/zenodo.3961647}{doi:10.5281/zenodo.3961647}.

\appendix
\section{Helical decay in 2.5-D}
\label{HelicalDecay}

\begin{figure}\begin{center}
\includegraphics[width=\columnwidth]{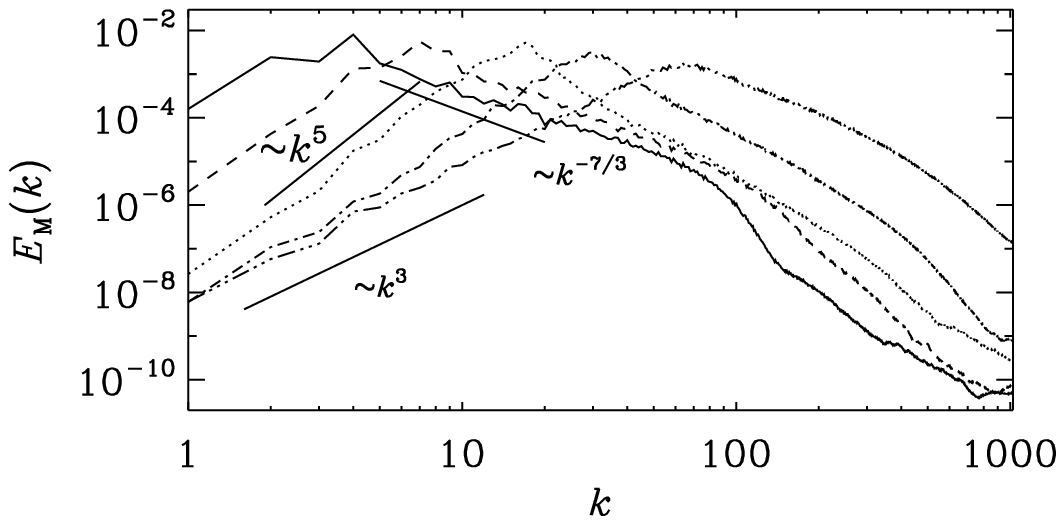}
\end{center}\caption[]{
Spectral evolution for Run~N with a helical magnetic field
at times $t=0.001$, 0.1, 1, 10, and 80 for $k_0=60$ and $A_0=1$.
}\label{ppower_comp_hel}\end{figure}

A 2.5-D magnetic field can be fully helical.
In \Fig{ppower_comp_hel}, we show the magnetic energy spectra for Run~N.
Note that the height of the peak increases slightly with time.
This might be a consequence of the fact that at late times, the spectra have a more distinct and sharper peak.

\section*{Authors' ORCIDs}

\noindent
A. Brandenburg, https://orcid.org/0000-0002-7304-021X

\noindent
V. Kumar, https://orcid.org/0009-0008-2158-3774

\bibliography{ref}{}
\bibliographystyle{jpp}
\end{document}